\documentclass[aps,prl,twocolumn,superscriptaddress,longbibliography,bibnotes]{revtex4-2}
\usepackage[utf8]{inputenc}
\usepackage{diagbox}
\usepackage{diagbox}
\usepackage{amsmath}
\usepackage{amsfonts}
\usepackage{physics}
\usepackage[space]{grffile}
\usepackage{graphicx} 
\usepackage{amssymb}
\usepackage{upgreek}
\usepackage{lineno}
\usepackage{hyperref}

\usepackage{natbib}
\usepackage{braket}
\usepackage{float}
\usepackage{color}
\usepackage{blindtext}
\usepackage{comment}
\usepackage{graphicx}
\usepackage{xr-hyper}
\usepackage{hyperref}
\usepackage{xcolor}
\hypersetup{colorlinks,breaklinks,
            urlcolor=[rgb]{0,0,0.64},
            linkcolor=[rgb]{0,0,0.64},
            citecolor=[rgb]{0,0,0.64},
            filecolor=[rgb]{0,0,0.64}}
\usepackage[T1]{fontenc}
\usepackage{cleveref}

\begin{document}
\title{Sublattice-resolved coherent phonon dynamics in charge density waves}

\author{Kyoung Hun Oh}
\thanks{These authors contributed equally to this work}
\affiliation{Department of Physics, Massachusetts Institute of Technology, Cambridge, MA 02139, USA}

\author{Honglie Ning}
\thanks{These authors contributed equally to this work}
\affiliation{Department of Physics, Massachusetts Institute of Technology, Cambridge, MA 02139, USA}

\author{Zongqi Shen}
\thanks{These authors contributed equally to this work}
\affiliation{Department of Physics, Massachusetts Institute of Technology, Cambridge, MA 02139, USA}

\author{Yifan Su}
\thanks{These authors contributed equally to this work}
\affiliation{Department of Physics, Massachusetts Institute of Technology, Cambridge, MA 02139, USA}

\author{Jack Maier}
\affiliation{Department of Physics, Massachusetts Institute of Technology, Cambridge, MA 02139, USA}

\author{Gyeongbo Kang}
\affiliation{PAL-XFEL, Pohang Accelerator Laboratory, Pohang, Gyeongbuk 37673, Republic of Korea}

\author{Hyeongi Choi}
\affiliation{PAL-XFEL, Pohang Accelerator Laboratory, Pohang, Gyeongbuk 37673, Republic of Korea}

\author{Dong Wu}
\affiliation{Beijing Academy of Quantum Information Sciences, Beijing 100913, China}

\author{Qiaomei Liu}
\affiliation{International Center for Quantum Materials, School of Physics, Peking University, Beijing 100871, China}

\author{Hyun-Woo J. Kim}
\affiliation{Department of Physics, Pohang University of Science and Technology, Pohang 37673, Republic of Korea}

\author{Seunghyeok Ha}
\affiliation{Department of Physics, Pohang University of Science and Technology, Pohang 37673, Republic of Korea}

\author{Jaehwon Kim}
\affiliation{Department of Physics, Pohang University of Science and Technology, Pohang 37673, Republic of Korea}

\author{Byungjune Lee}
\affiliation{Department of Physics, Pohang University of Science and Technology, Pohang 37673, Republic of Korea}
\affiliation{Max Planck POSTECH/Korea Research Initiative, Center for Complex Phase of Materials, Pohang 37673, Republic of Korea}

\author{B. J. Kim}
\affiliation{Department of Physics, Pohang University of Science and Technology, Pohang 37673, Republic of Korea}

\author{N. L. Wang}
\affiliation{Tsung-Dao Lee Institute, Shanghai Jiao Tong University, Shanghai 200240, China}

\author{Yao Wang}
\affiliation{Department of Chemistry, Emory University, Atlanta, GA 30322, USA}

\author{Hoyoung Jang}
\affiliation{PAL-XFEL, Pohang Accelerator Laboratory, Pohang, Gyeongbuk 37673, Republic of Korea}
\affiliation{Photon Science Center, Pohang University of Science and Technology, Pohang 37673, Republic of Korea}

\author{Nuh Gedik}
\email[email: ]{gedik@mit.edu}
\affiliation{Department of Physics, Massachusetts Institute of Technology, Cambridge, MA 02139, USA}

\begin{abstract}
Phonons govern fundamental material properties and play a central role in various electronic phase transitions. 
Coherent driving of specific phonon modes enables on-demand phase control, motivating sublattice-resolved identification of real-space phonon motions. 
Yet experimentally resolving these motions remains challenging, limiting precise phonon-based control. 
Here, we introduce a dynamical protocol to track element-resolved phonon dynamics in the charge density wave material EuTe$_4$, in which the dominant Te-sublattice charge order is accompanied by a previously unreported Eu-sublattice component.
We leverage the elemental selectivity of time-resolved resonant X-ray scattering to reveal three coherent phonon modes with distinct sublattice character, thereby disentangling Eu- and Te-dominated lattice dynamics, in good agreement with theoretical calculations of the phonon eigenvectors.
This time-domain approach, which surpasses the energy-resolution limits of conventional frequency-domain inelastic scattering, provides a broadly applicable framework for decomposing coherent phonons in multi-element materials, which is crucial for the targeted control of phases of matter.

\end{abstract}
\maketitle

Phonons not only govern the thermal and mechanical properties of materials but also play a critical role in shaping their electronic phases \cite{Ashcroft1976SolidPhysics}.
Recent advances have shown that selective phonon excitation can dynamically stabilize exotic quantum phases, such as superconductivity \cite{Fausti2011Light-inducedCuprate,Mankowsky2014NonlinearYBa2Cu3O6.5}, magnetism \cite{Disa2020PolarizingField}, and ferroelectricity \cite{Nova2019Metastablesub3/sub, Disa2021EngineeringLight}.
In these cases, driving specific vibrational modes induces collective atomic displacements that reshape the orders through electron–phonon coupling.
A central prerequisite for such targeted control is to identify which sublattices are predominantly involved in a given phonon.

Experimental techniques to directly probe phonons include Raman/infrared spectroscopy \cite{Albrecht1961OnIntensities, DeFaria1997RamanOxyhydroxides, Kuo1982InfraredHydrogen}, Brillouin scattering \cite{Merklein2022100Perspectives}, transient optical reflectivity \cite{Hase1998DynamicsPulses, Qi2010UltrafastCrystals}, and inelastic X-ray/neutron/electron scattering \cite{Ibach1970OpticalSpectroscopy, Mittal2006ModelingScattering,Burkel1987ObservationPhonons}, and time-resolved electron or X-ray elastic scattering.
While these methods provide valuable access to phonon energies, lifetimes, and dispersions, they generally do not reveal the atomic displacement patterns associated with phonon eigenvectors.
First, most conventional optical probes and non-resonant scattering techniques lack elemental sensitivity, preventing direct determination of how individual atomic species participate in a given vibrational mode. 
Second, while conventional non-resonant time-resolved scattering can, in principle, determine phonon eigenvectors from the temporal evolution of diffraction patterns, doing so requires tracking numerous Bragg reflections, making such experiments experimentally demanding and often impractical. 
Finally, although resonant inelastic X-ray scattering (RIXS) provides elemental selectivity, its energy resolution fundamentally limits its ability to resolve soft phonons in the sub-10-THz regime.
Consequently, sublattice-resolved lattice vibrations are typically inferred by complementing experimental probes with supporting density functional theory (DFT) calculations.
However, for complex compounds with large unit cells, including superstructures and artificial van der Waals heterostructures, this workflow becomes computationally demanding, and phonon-mode identification can become increasingly ambiguous, particularly when many modes cluster in frequency.
This uncertainty not only hampers the determination of phonon eigenvectors in complex materials, but also limits efforts to engineer and manipulate electronic phases at the microscopic level.

\begin{figure*}[t]
\includegraphics[width=6.75in]{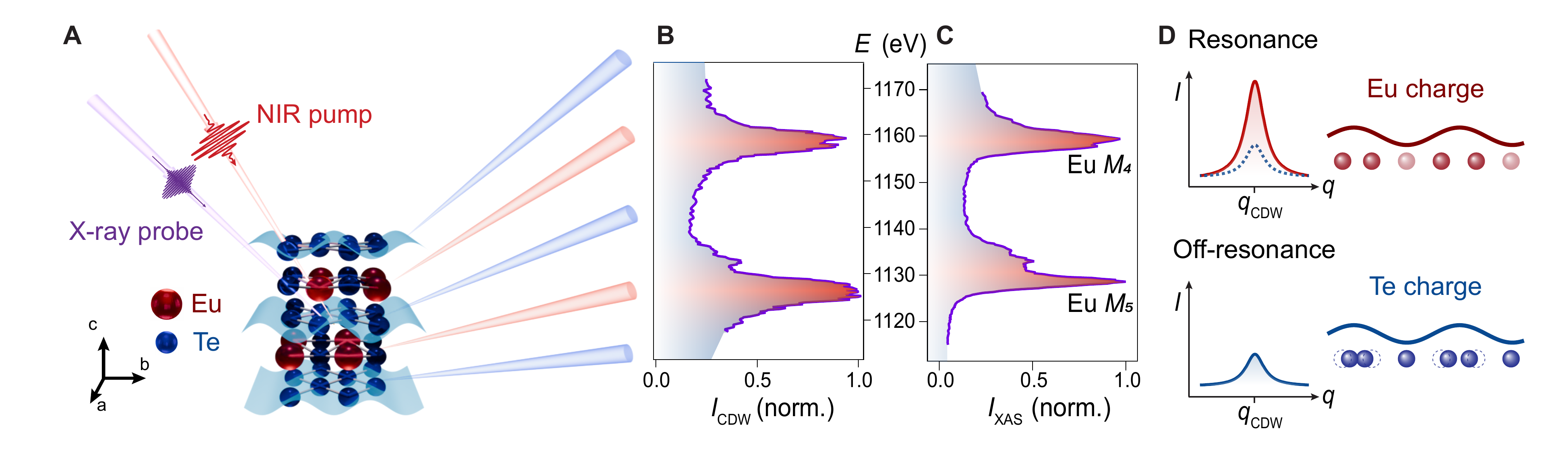}
\caption{Illustration of the experimental scheme.
(\textbf{A}) Schematic of the time-resolved resonant X-ray scattering (tr-RXS) setup, employing a near-infrared pump centered at 800 nm with a fluence of $F = 4$ mJ/cm$^2$ and a soft X-ray probe. Eu and Te atoms are shown in red and blue, respectively. The charge-lattice order in the Te layers are highlighted by blue waveforms. Diffracted X-rays with energies near the Eu $M_5$ (or $M_4$) resonance edge are shown in red, while those far from the resonance are shown in blue. 
(\textbf{B}) X-ray photon energy dependent intensity of the monolayer CDW peak at fixed (0 0.356 2).
(\textbf{C}) Fluorescence yield XAS spectrum of EuTe$_4$, exhibiting two absorption peaks (Eu $M_5$: 1128 eV and $M_4$: 1159 eV). In (b,c), red and blue shaded regions correspond to X-ray probe energies near and far from the Eu $M$ resonance edges, respectively.
(\textbf{D}) Schematic illustrating the resonant enhancement of CDW peak intensity. Blue and red peaks correspond to off-resonant and resonant X-ray probe cases (left panel) that describe the Te and Eu charge-order components (right panel), respectively.
}
\label{Fig1}
\end{figure*}

To overcome this challenge, an approach to track the sublattice-specific lattice dynamics with elemental sensitivity is required.
Time-resolved resonant X-ray scattering (tr-RXS) uniquely offers this capability \cite{Fink2013ResonantScattering, Lee2012PhaseNickelate, Chuang2013Real-timeDiffraction, Moore2016Ultrafast3, Rettig2016ItinerantHo, Rettig2019DisentanglingDiffraction, Windsor2022ExchangeAntiferromagnets, Jang20234DLaser, Jiang2025UsingMaterial}. 
While tr-RXS has previously been employed to disentangle electronic and structural order parameters in manganites \cite{Rettig2019DisentanglingDiffraction}, its use for experimentally resolving the atomic-sublattice character of coherent phonons has remained largely unexplored.
Here, we show that, by tuning the X-ray energy near the absorption edge of a selected element, tr-RXS enhances sensitivity to specific atomic orbitals, enabling direct, sublattice-resolved probing of coherent phonon dynamics triggered by optical excitation (Fig.~\ref{Fig1}A,D).
This Fourier-transform-based time-domain approach allows access to modes across a broad momentum range and frequencies spanning the THz and even GHz regimes, surpassing the energy resolution limits of RIXS \cite{Albrecht1961OnIntensities, DeFaria1997RamanOxyhydroxides,Ament2011ResonantExcitations, deGroot2024ResonantScattering}.

A particularly relevant context for applying this experimental scheme is provided by charge density wave (CDW) materials, as they frequently accommodate multiple atomic species, while the CDW develops on a specific sublattice and is accompanied by lattice distortions arising from strong charge-lattice coupling.
Probing sublattice-specific phonon dynamics can thus elucidate which sublattices participate in the CDW distortion and which phonon eigenvectors govern the CDW modulation.
To this end, we choose the newly synthesized bi-elemental CDW material EuTe$_4$ as a testbed (SI appendix, Supplemental Note 1), which comprises alternating Te monolayer and Te bilayer square sheets separated by EuTe slabs along the $c$-axis (Fig.~\ref{Fig1}A) \cite{Wu2019LayeredSheets, Lv2022UnconventionalWave, Rathore2023EvolutionEuTe4, Zhang2023ThermalEuTe4}.
Previous studies have revealed periodic charge and lattice distortions exclusively within the Te monolayers, characterized by an incommensurate CDW wavevector $\mathbf{q} = (0, q_b, 0)$ with $q_b = 0.644$ \cite{Wu2019LayeredSheets, Lv2025LargeWaves,Xiao2024HiddenEuTe4,Lv2025LargeWaves}, whereas the EuTe slabs do not exhibit a clear lattice distortion \cite{Wu2019LayeredSheets, Lv2022UnconventionalWave, Rathore2023EvolutionEuTe4, Xiao2024HiddenEuTe4, Lv2025LargeWaves}. 
Nevertheless, electrostatic interactions can potentially imprint the charge modulation of the Te layers onto the EuTe slabs, inducing an additional charge modulation on the Eu sites with the same periodicity, $\mathbf{q}$ \cite{Lee2012ResonantTritellurides}. 
Now, by performing tr-RXS to probe the CDW superlattice diffraction peak while tuning the incident X-ray energy on and off the Eu absorption edge, we can selectively couple to the Eu and Te components of the CDW, respectively. 
This energy-dependent measurement thus allows independent, layer-by-layer determination of the Te and Eu sublattice dynamics and enables direct comparison of their respective phonon dynamics.

We first examine the possible existence of contribution of the Eu-sublattice to the CDW, by measuring the equilibrium CDW intensity using RXS across a range of probe energies (Fig.~\ref{Fig1}B) \cite{Lee2012ResonantTritellurides,Li2022DiscoveryCsV3Sb5}.
When the photon energy is detuned sufficiently far (>$\sim$10 eV) from any elemental resonances, the (0 $1-q_b$ 2) superlattice scattering intensity is dominated by the Te charge–lattice–coupled spatial distortion (``Te charge'', Fig.~\ref{Fig1}D, see SI appendix, Supplemental Note 2) \cite{Fink2013ResonantScattering,Wu2019LayeredSheets}. 
By contrast, when the incident energy is tuned to Eu $M$ edges, whose positions are precisely determined by simultaneous X-ray absorption spectroscopy (XAS) (Fig.~\ref{Fig1}C), the superlattice intensity exhibits a pronounced resonant enhancement.
We note that this resonant enhancement does not originate from an increase in the uniform fluorescence background (Supplemental Note 3).
Since the Eu elemental sensitivity is significantly amplified under these conditions, this enhancement confirms the Eu-sublattice contribution to the CDW (``Eu charge'', Fig.~\ref{Fig1}D).
Such a contribution likely arises from additional charge modulation on the Eu sublattice with the same CDW wavevector $\mathbf{q}$, rather than solely from hybridization between the Eu 4$f$ and Te 5$p$ orbitals, as these states are both spatially and energetically well separated owing to the highly localized nature of the Eu 4$f$ orbitals \cite{Wu2019LayeredSheets}, and no clear transient change in the Eu $M$-edge XAS is observed upon photoexcitation (Supplemental Note 4).

\begin{figure*}[t]
\includegraphics[width=6.75in]{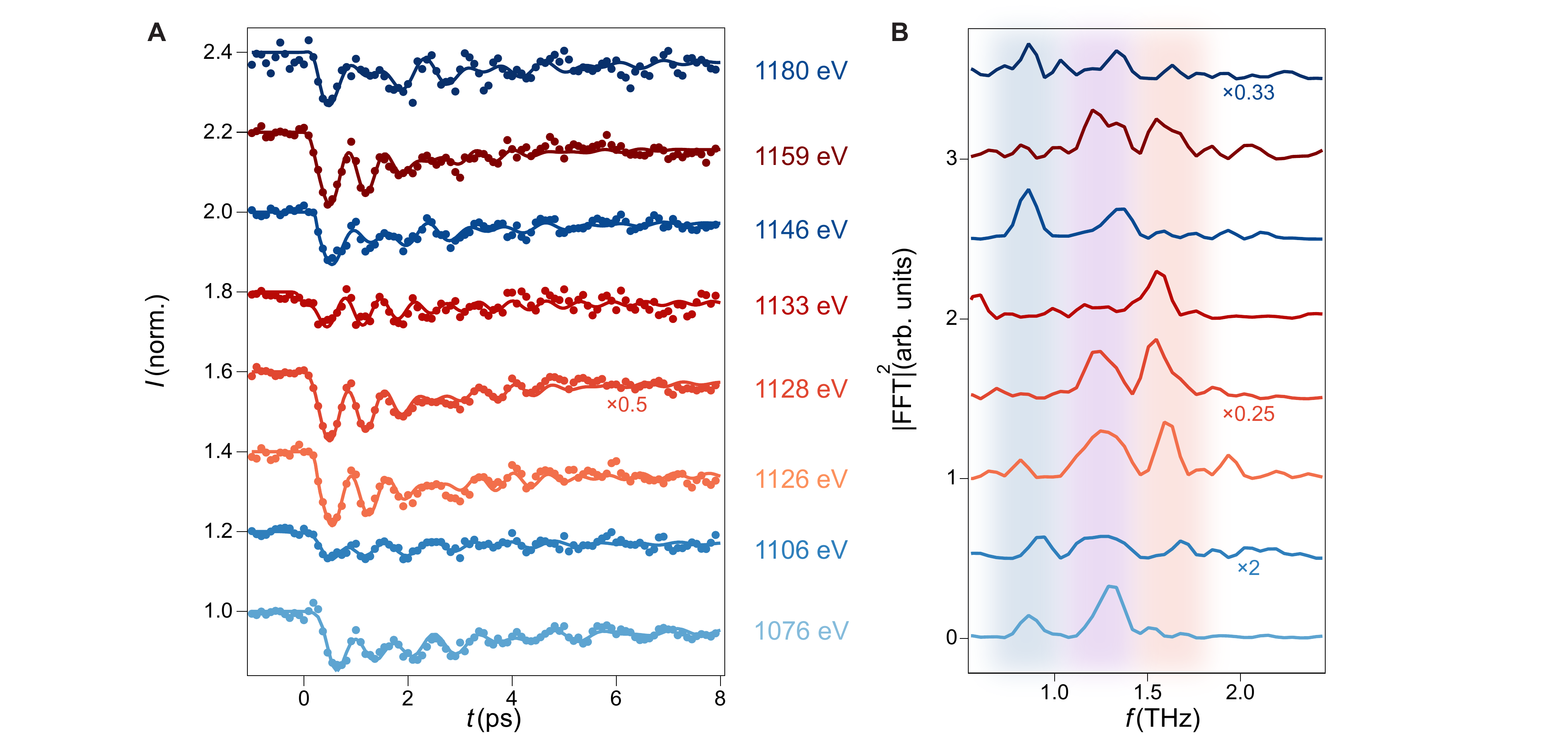}
\caption{Energy-dependent time-resolved resonant X-ray scattering (Tr-RXS) results. 
(\textbf{A}) Temporal evolution of the normalized intensity of the monolayer CDW peak at (0 0.356 2) to their equilibrium values measured with different probe energies. Each time trace is vertically offset for visual clarity. Time traces measured with probe energies near the Eu $M$ resonance edges are shown in reddish color, while those measured far from resonance are shown in bluish color. Solid lines are fits to damped oscillators with an exponential decaying background. 
(\textbf{B}) Fast Fourier transform (FFT) of the oscillatory part of the time traces shown in (\textbf{A}). Each curve is vertically offset for clarity. Blue, purple and red shading represents the phonon modes at $\sim$0.85 THz, $\sim$1.3 THz, and $\sim$1.6 THz, respectively.}
\label{Fig2}
\end{figure*}

\begin{figure*}[t]
\includegraphics[width=6.75in]{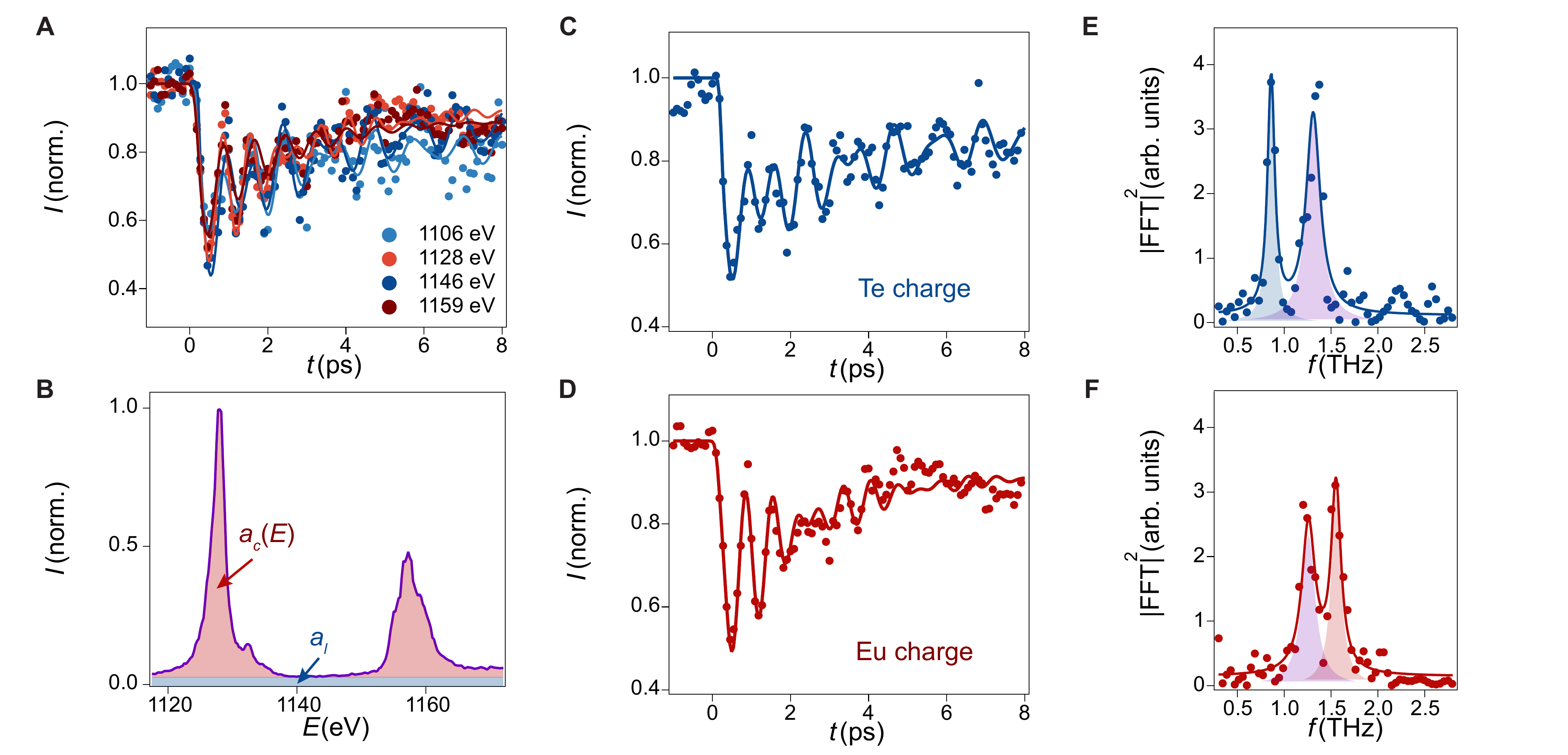}
\caption{Quantitatively disentangling the Te and Eu charge components dynamics.
(\textbf{A}) Temporal evolution of the intensity normalized to their equilibrium values for four representative probe energies with variations of the probe penetration depth corrected. Solid lines are fits to damped oscillators atop an exponential decaying background.
(\textbf{B}) Energy-dependent CDW peak intensity with variations of the probe penetration depth corrected. Blue and red shaded area represent the intensity contributed from the Te CDW component ($a_{l}$) and Eu charge order ($a_{c}(E)$), respectively. $a_{l}$ is assumed to be a constant within the investigated X-ray energy range.
(\textbf{C},\textbf{D}) Decoupled temporal evolution of the Te and Eu contribution to the CDW diffraction intensity, respectively. Solid lines are fits to damped oscillators atop an exponential decaying background.
(\textbf{E},\textbf{F}) FFT of the oscillatory components of the time traces in (\textbf{C},\textbf{D}). Blue, purple and red shading represents the phonon modes at $\sim$0.85 THz, $\sim$1.3 THz, and $\sim$1.6 THz, respectively. We note that the peaks near 1.3 THz in (\textbf{E},\textbf{F}) differ by $\sim$0.1 THz, which may reflect the presence of multiple closely spaced phonon modes. As this possibility does not affect the main conclusions of this work, we treat this feature as a single phonon mode for simplicity.}
\label{Fig3}
\end{figure*}

We apply the aforementioned tr-RXS-based strategy to track the temporal evolution of the CDW diffraction intensity at (0 $1-q_b$ 2), following excitation by an 800 nm pump pulse (Fig.~\ref{Fig1}A), while varying the probe energy across the Eu $M$ absorption edges (Fig.~\ref{Fig2}A, SI appendix, Supplemental Note 2).
As expected for typical photoinduced CDW melting \cite{Lee2012PhaseNickelate, Zong2019EvidenceTransition, Ning2024DynamicalSuperconductor, Ning2025BidirectionalCompetition, Oh2025JointDefects}, all time traces exhibit a reduction within $\sim$0.5 ps without any indication of peak shift (SI appendix, Supplemental Note 5), followed by a recovery toward a quasi-equilibrium value over $\sim$6 ps.
Superimposed on this melting–recovery behavior are oscillations in diffraction intensity, which we attribute to coherent phonons launched by pump excitation (SI appendix, Supplemental Note 6-8) \cite{Huber2014CoherentTransition, Singer2016PhotoinducedAmplitude, Moore2016Ultrafast3, Trigo2019CoherentSmTe3, Nguyen2023UltrafastI, Ning2024DynamicalSuperconductor}, without alteration of the average lattice structure (SI appendix, Supplemental Note 9).

Fourier transform of the oscillatory components in the tr-RXS traces reveals distinct modes at around 0.85 THz, 1.3 THz, and 1.6 THz, whose visibility strongly depends on the probe energy (Fig.~\ref{Fig2}B).
A clear contrast arises between resonant and off-resonant conditions: when the probe energy is tuned near the Eu $M$ edges, the 1.3 THz and 1.6 THz oscillations dominate, whereas off-resonant probing highlights the 0.85 THz and 1.3 THz modes.
Since X-rays tuned to different energies provide selective sensitivity to the CDW on the Te versus Eu sublattices, these results indicate that the 0.85 THz mode predominantly involves Te atomic motion, the 1.6 THz mode primarily involves Eu motion, and the 1.3 THz mode involves contributions from both sublattices. 
This qualitative conclusion unambiguously demonstrates the selective sensitivity to atomic motions of both Te monolayers and EuTe spacer layers using our tr-RXS approach.

\paragraph{}
\begin{figure}[t]
\includegraphics[width=3.375in]{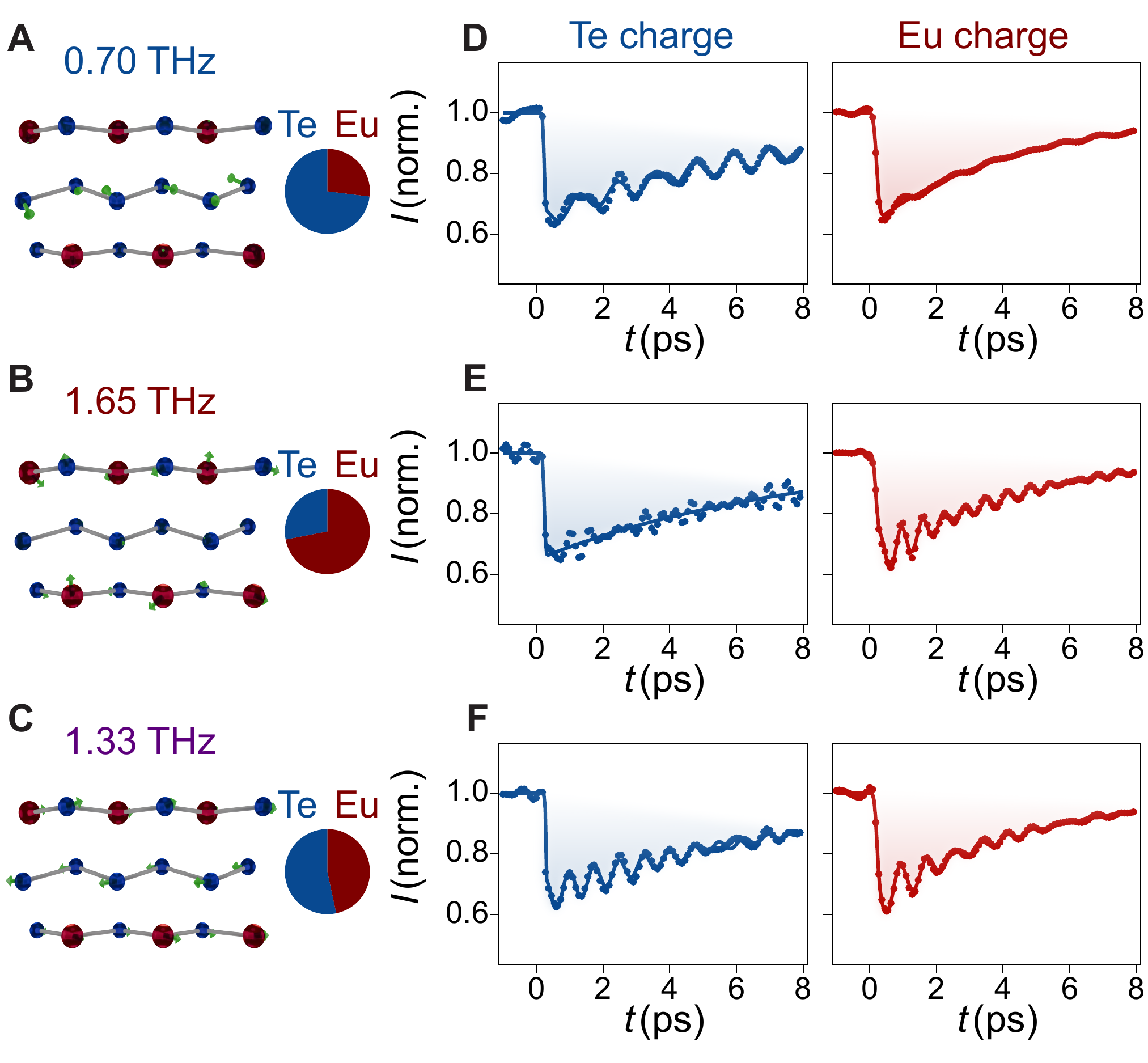}
\caption{Phonon eigenvectors and a physical picture accounting for the selective appearance of the Te-only mode.
(\textbf{A}) Schematic of the eigenvector of the phonon at 0.70 THz, which predominantly involves the motion of Te atoms in the Te monolayers. 
(\textbf{B}) Schematic of the eigenvector of the phonon  at 1.65 THz, which predominantly involves the motion of  Eu atoms in the EuTe layers. 
(\textbf{C}) Schematic of the eigenvector of the phonon  at 1.33 THz, which predominantly involves both the motion of Te atoms in the Te monolayers and Eu atoms in the EuTe layers. Green arrows depict their eigenvectors. Red and blue balls are Eu and Te atoms, respectively. Bilayer Te layers are omitted in the structures for visual clarity. Pie charts on the right display the ratio of the average atomic displacements for Te atoms in the Te monolayers and Eu atoms in the EuTe layers for each phonon.
(\textbf{D}-\textbf{F}) Temporal evolution of the Te and Eu contribution to the CDW diffraction
intensity in Figs.~\ref{Fig3}(\textbf{C},\textbf{D}) with band-pass filtering around the three phonon frequencies shown in (\textbf{A}-\textbf{C}). Solid lines are fits to a single damped oscillator at the corresponding phonon frequency atop an exponential decaying background. 
We note that discrepancies in frequency exist between the experimentally measured and DFT-calculated phonon modes. These differences arise because multiple phonon modes clustered within a narrow frequency range can possess similar eigenvectors. Consequently, the phonon eigenvectors shown here should be regarded as representative rather than unique assignment.
}
\label{Fig4}
\end{figure}

To quantitatively disentangle the dynamics of the Te and Eu charge components of the order, we construct a model that leverages all tr-RXS traces taken at various photon energies (SI appendix, Supplemental Note 2).
We first normalize the raw tr-RXS intensity time traces by accounting for variations in X-ray probe penetration depth as a function of probe energy, which enables direct comparison of CDW peak suppression and oscillation amplitudes across different energies.
The corrected time traces (Fig.~\ref{Fig3}A), which reflect the response of the topmost crystal layer, collapse onto a single curve, indicating that the Te and Eu charge components exhibit a similar degree of melting upon photoexcitation (SI appendix, Supplemental Note 2).
The normalized CDW peak intensity $I(E,t)$ measured at different X-ray energies $E$ can then be treated as a weighted sum of contributions from the two sublattices \cite{Fink2013ResonantScattering, Achkar2013ResonantSuperconductors, Takubo2014BondIrTe2}:
\begin{equation}\label{EuTe}
\begin{split}
    I (E, t) = \frac{a_{l} I_{\textrm{Te}}(t)+a_{c}(E) I_{\textrm{Eu}}(t)}{a_{l}+a_{c}(E)}.
\end{split}
\end{equation}
Here, $I_{\textrm{Te}}(t)$ and $I_{\textrm{Eu}}(t)$ denote the normalized temporal evolution of the Te and Eu charge order components, and $a_{l}$ and $a_{c}(E)$ represent their respective weights. 
We assume $a_{l}$ independent of $E$ and $a_{c}$ strongly dependent on $E$, with their values determined from the energy-dependent CDW diffraction intensity (Fig.~\ref{Fig3}B).
This framework allows us to precisely extract the sublattice dynamics $I_{\textrm{Te}}(t)$ and $I_{\textrm{Eu}}(t)$ by globally fitting $I (E, t)$ traces collected at multiple probe energies.
The resulting decoupled time traces clearly reveal oscillatory components with distinct frequencies (Figs.~\ref{Fig3}C,D).
Their corresponding FFT spectra show that the Eu contribution features oscillations at 1.3 THz and 1.6 THz, while the Te contribution exhibits modes near 0.85 THz and 1.3 THz (Figs.~\ref{Fig3}E,F), consistent with the conclusion drawn from the raw data (Fig.~\ref{Fig2}B).
Moreover, to isolate mode-specific contributions to Eu and Te components, we band-pass-filter the decoupled time traces at the three phonon frequencies (Figs.~\ref{Fig4}D-F), which confirms our assignment of phonon structures and highlighting the sublattice selectivity of our method.

To further corroborate the sublattice character of the three phonon modes, we perform first-principles DFT calculations \cite{Perdew1996GeneralizedSimple} and identify phonon modes with frequencies close to the experimental values that exhibit atomic motions in good agreement with the experimental indications (Fig.~\ref{Fig4}A-C, SI appendix, Supplemental Note 10):
a 0.7 THz mode predominantly involving Te displacements in Te monolayers (Fig.~\ref{Fig4}A), consistent with its role in modulating the Te CDW (Fig.~\ref{Fig4}D), a 1.65 THz phonon dominated by Eu motions in the EuTe spacer layer (Fig.~\ref{Fig4}B), thus emerging when the X-ray energy is resonant with Eu edges (Fig.~\ref{Fig4}E), and a 1.33 THz mode comprising both Eu and Te distortions (Fig.~\ref{Fig4}C), consistent with the mixed character inferred from the experiment (Fig. \ref{Fig4}F). 

In conclusion, our energy-dependent tr-RXS measurements uncover distinct coherent oscillations of the Te monolayer and the EuTe space layer, establishing sublattice-specific, layer-by-layer phonon dynamics in EuTe$_4$. 
These results demonstrate the unique capability of tr-RXS to resolve photoinduced atomic motions with element-level specificity.
This approach is well suited for resolving sublattice-resolved phonon dynamics in multi-element systems with large unit cells, including CDW superstructures and artificial van der Waals heterostructures.
By revealing the sublattice-specific character of individual phonons, this work lays the foundation for selectively and resonantly driving targeted phonons using terahertz pulses, advancing the burgeoning field of on-demand phononic control of material properties.


\section*{Acknowledgments}
The authors thank Riccardo Comin, Alberto de la Torre,  Hoon Kim, Mingu Kang, Batyr Ilyas, Alfred Zong, and B. Q. Lv for fruitful discussions. The work at MIT was supported by the U.S. Department of Energy, the BES DMSE (data collection and analysis) and the Gordon and Betty Moore Foundation's EPiQS Initiative grant GBMF9459 (manuscript writing). N.L.W acknowledges support from National Natural Science Foundation of China (Grant No. 12488201). N.L.W and D.W are supported by National Key Research and Development Program of China (2024YFA1408700). B.L. and H.J. acknowledges support by the National Research Foundation grant funded by the Korea government (MSIT) (Grant No. RS-2022-NR068223). The tr-RXS experiments were performed at the SSS-RSXS end station (Proposal No. 2024-1st-SSS-008 and 2024-2nd-SSS-008) of the PAL-XFEL funded by the Korea government (MSIT). Data and network services at PAL-XFEL were supported by GSDC and KREONET, provided by the Korea Institute of Science and Technology Information (KISTI). 

\bibliography{references_arxiv.bib}
\end{document}